\documentclass[aps,psfig,prl,showpacs,                   twocolumn]{revtex4}
\usepackage{amssymb}

\usepackage{dcolumn}
\usepackage{amsmath}
\usepackage{graphicx}
\usepackage{latexsym}

\begin{document}

\title{Asymptotics of the dispersion interaction:\ analytic benchmarks for van
der Waals energy functionals}
\date{Original receipt date Feb 20 2005, revised MS July, August, Nov 05}
\author{John F. Dobson$^{1,2,3}$, Angela White$^{1}$, and Angel Rubio$^{4,3}$}

\begin{abstract}
We show that the usual sum of $R^{-6}$ contributions from elements separated
by distance $R$ can give \emph{qualitatively} wrong results for the
electromagnetically non-retarded van der Waals interaction between
non-overlapping bodies. This occurs for anisotropic nanostructures that have
a zero electronic energy gap, such as metallic nanotubes or nanowires, and
nano-layered systems including metals and graphene planes.
In all these cases
our analytic microscopic calculations give an interaction falling off
with a power of separation different from the conventional value. We discuss
implications for van der Waals energy functionals.  The new nanotube
interaction might be directly observable at sub-micron separations.
\end{abstract}


\affiliation{$^1$Nanoscale Science and Technology Centre 
, Griffith University, Nathan, Queensland 4111, Australia} 

\affiliation{$^2$LSI, Ecole Polytechnique, F91128 Palaiseau, France} 

\affiliation{$^3$Donostia International Physics 
Center (DIPC), E20018 Donostia/San Sebasti\'{a}n, Spain}

\affiliation{$^4$
European Theoretical Spectroscopy Facility (ETSF) and 
Departamento de F\'{i}sica de Materiales,
UPV/EHU and Unidad de Materiales Centro Mixto
CSIC-UPV/EHU San Sebasti\'{a}n, Spain}

\maketitle






Dispersion interactions (part of the van der Waals, vdW energy)\cite
{vdWBooks2} are especially significant in soft matter. The vdW physics that
we expose here could be
relevant in predicting the
energetics of
bundles of metallic nanowires or nanotubes, layered metallic systems, $\pi -$%
conjugated systems including graphite, intercalated graphite, graphitic
hydrogen storage systems and pi-stacked biomolecules, and other weakly bound
(''soft'') layered and striated nanosystems. Standard local (LDA) and
gradient (GGA) 
density functionals for the electronic energy do not obtain any distant
dispersion interaction, but density functionals have been derived recently
that obtain, in a natural fashion, both distant dispersion interactions and
their saturation at small distances. These and other numerically practicable
vdW energy schemes available to date 
\cite
{vdWBooks2,LifshitzMolAttrFrcsSolsJETP56,UnivGraphiticPotlGirifalco,HardMNosSoftMattRydbergEtal,TractableNonlocVdWSlabs,HardNosSoftMattRevisRydberg+,vdWFnalGenGeomDionPRL04,CHARMMPotls98}
for the above systems (in the
electromagnetically non-retarded regime) have a ``universal'' feature: the
distant vdW interaction energy between sufficiently separated subsystems is
given qualitatively by a sum of contributions of form $R^{-6}$ between
microscopic elements separated by distance $R$. This leads to ``standard''
power laws $E\propto -D^{-p}$ for the interaction energy between various
macroscopic bodies separated by distance $D$ (column 3 of Table 1). Although
these ''universal'' asymptotic results are indeed valid for most macroscopic
systems, we show below that they fail for the anisotropic nanostructures
mentioned above. 
Column 2 of Table 1 summarizes the asymptotic ($D\rightarrow \infty $)
benchmarks that we propose below for the vdW energy of two parallel
nanostructures of infinite extent.

\begin{table}[tbp]
\begin{tabular}{lll}
System & present & standard \\ 
&  &  \\ \hline
&  &  \\ 
1D metals {*} & $\mathbf{-D^{-2}\left( \ln (KD)\right) ^{-3/2}}$ & $-D^{-5}$
\\ 
1D insulators \cite{JFDAWhiteUnpub} & $-D^{-5}$ & $-D^{-5}$ \\ 
2D metals\cite{FractvdWThinMetFilmsBostromSernelius,JFDEtAlAustJChem02} & $%
\mathbf{-D^{-5/2}}$ & $-D^{-4}$ \\ 
$\pi $-conjugated layers {*} & $\mathbf{-D^{-3}}$ & $-D^{-4}$ \\ 
1 metallic, 1 $\pi $-layer {*} & $\mathbf{-D^{-3}\ln (D/D_0)}$ & $-D^{-4}$
\\ 
2D insulators\cite{HardNosSoftMattRevisRydberg+} & $-D^{-4}$ & $-D^{-4}$ \\ 
\emph{Thick} metals or ins.\cite{JFDEtAlAustJChem02} & $-D^{-2}$ & $-D^{-2}$%
\end{tabular}
\caption{Asymptotic vdW energy of parallel structures. $K$ and $D_0$ are
constants. * denotes new derivations given here.}
\end{table}

To analyze these situations, we use the correlation energy $E_c^{RPA}(D)$
from the Random Phase Approximation (RPA)\cite
{jfdvdw,SurfEnPitarkeEguiluz,RPAMolecsFurche,RPAMolecsFuchsGonze02,DobsonWangPRL99,JeilJFDPabloRex04}%
, a basic microscopic theory\ that does not rely on assumptions of locality,
additivity nor $R^{-6}$ contributions.
Going beyond the RPA
 does not change the asymptotic power laws predicted here,
unless the exchange-correlation kernel $f_{xc}$
\cite{GrossKohnPRL85,EnOptFxc,JeilJFDPabloRex04} has a slower spatial
decay than the bare coulomb interaction, an unprecedented and unlikely
scenario.

 Where the separated subsystems exhibit lightly
damped long-wavelength plasmons, we note\cite{JFDIJQC04} that the principal
contribution to $E_c^{RPA}(D)$ comes from the sum of coupled-plasmon
zero-point energies:\ otherwise we use the full RPA. Some essential common
features of these systems will be abstracted from these specific
calculations. We obtain analytic results for the asymptotic $(D\rightarrow
\infty )\,$regime in all cases, but in section E we will also discuss
systems near their equilibrium spacing.


\emph{A: Distant attraction between metallic linear structures.} Consider
two parallel, infinitely long conducting wires or tubes separated by a
distance $D$ substantially exceeding their radius $b$, and with $b<\lambda $
where $\lambda $ is a bulk screening length. Both standard $\sum R^{-6}\,$%
analysis\cite{UnivGraphiticPotlGirifalco} based on the vdW interaction
between electrons localized in atoms or bonds, and recent functionals \cite
{vdWInterParPolymersNanotubes+Hyldgaard+05}, give a vdW energy per unit
length of the form $E\propto -D^{-5}$. Instead we consider the zero-point
energy of the delocalized coupled one-dimensional plasmon modes with
wavenumber $q\,$parallel to the long axis
\cite{CollectiveExcMetalNanotubesLongeBose93,JFDAWhiteUnpub}.%
The radially-smeared intra-wire coulomb interaction is $%
w_{11}(q)=w_{22}(q)=-2e^2\ln (qb)$ where 1 and 2 refer to the two wires, and
we have assumed $qb<<1$, as appropriate when $D>>b$. In the same limit the
bare density-density response for electronic motion parallel to the wire is $%
\chi _{011}=\chi _{022}=N_0q^2/(m\omega ^2)$ where $\omega \,$is the
frequency, $N_0$ is the number of electrons per unit length and $m$ is the
electron mass. RPA screening yields the interacting response of a single
wire as $\chi _{11}=\chi _{011}/(1-w_{11}\chi _{011})$. 
The inter-wire coulomb interaction in the present limit has a Bessel form, $%
w_{12}=2e^2K_0(qD)$. The RPA equation for coupled 1D plasmons on two
identical wires is $\chi _{11}^2w_{12}^2=1$, giving two roots for each $q$: $%
\omega _{\pm }(D)=c_{1D}\left| q\right| (\left| \ln (qb)\right| \pm
K_0(qD))^{1/2}$. Here $c_{1D}=(2N_0e^2/m)^{1/2}$ is a characteristic
velocity. The vdW energy is the separation-dependent part of the sum of
zero-point plasmon
energies per unit length: 
\[
\frac{E^{vdW}}L=\frac 1{2\pi }\int_{-\infty }^\infty \frac \hbar 2\left(
\omega _{+}(D)+\omega _{-}(D)-2\omega _{+}(\infty )\right) dq 
\]

For $D>>b$ we expanded to 2nd order in $K_0(qD)/\left| ln(qb)\right|$ which is small near
the peak of the integrand. This gave \cite{JFDAWhiteUnpub} approximately
\begin{equation}
E^{vdW}/L\approx -(16\pi )^{-1}\hbar c_{1D}D^{-2}(\ln (2.39D/b))^{-3/2}.
\label{WireWirePlasmon}
\end{equation}
This approach is reasonable when the electron mean free path $d_0$ along the
wire satisfies $d_0>D$. In fact bismuth nanowires\cite
{PosThermopowerBiNanowireGrozav04} and conducting nanotubes 
\cite{PhysPropsNanotubesSaitoDresselh} can both have $d_0\ge 1$ micron. 
Eq. (\ref{WireWirePlasmon}) differs from the widely accepted result $E\propto
-D^{-5}$ by nearly three powers of $D$, and is necessarily dominant, at
sufficiently large $D,$ over any such higher-power contributions (arising from the
remaining bound sp$^2$ electrons (in nanotubes) and azimuthal $\pi $ plasmons). Our plasmon
model does give $D^{-5}$ if a pinning force is added to mimic an insulator 
\cite{JFDAWhiteUnpub}. 

\emph{B: Distant attraction between thin conducting layers} Consider infinite
parallel metallic plates separated by distance $D$ and of thickness $b$,
with $b<<D$, $b<\lambda$ where $\lambda$ is the bulk screening length. As is
already well-known
\cite{FractvdWThinMetFilmsBostromSernelius,JFDEtAlAustJChem02},
 the zero point energy of long-wavelength coupled 2D
plasmons leads to an attraction of form $E\propto -D^{-5/2}$. The $\sum
R^{-6}$ approach, correct for thin \emph{insulators}, gives $E\propto
-D^{-4} $, different by 1.5 powers of $D$.

\emph{C:\ Distant attraction between planar }$\pi $\emph{-conjugated systems.}
What
does the physics of long-wavelength excitations imply for the energetics of
layered planar $\pi$-conjugated systems, such as the controversial
\cite{BenedictMeasGraphiteLayerAttr,CohEnGraphiteThermDesorb+Hertel,OrientationC60sTournusCharlier05,GraphiteBindingSemiEmpirHasegawa+04}
and technologically important
\cite{PhysPropsNanotubesSaitoDresselh,HStorageCNanotubesGraphiticClustersSimonyan+}
graphene-based systems? Firstly, an isolated graphene layer at T=0K is not a
metal but a zero-gap insulator\cite{PhysPropsNanotubesSaitoDresselh}. Thus
one cannot argue for a metallic $-D^{-5/2}$ energetics (as under (B)\ above)
at large layer separation $D$ and $T=0K$, even though band overlap makes
graphite weakly metallic at the equilibrium layer spacing. We briefly derive
below, however, our new result that the attractive energy between two
well-separated graphene planes at $T=0K$ is of form $-C_3D^{-3}$, closer to
metallic $D^{-5/2}$ behavior than to insulating $D^{-4}$ behavior. All the
new physics here comes from electrons close to the Fermi level: we can
ignore the response of the tightly-bound covalent sp$^2$ electrons, whose
finite energy gap ensures that they produce a conventional vdW attraction of
2D insulator type (energy $\propto -D^{-4}$), negligible at large separations
compared with the $D^{-3}$ vdW attraction that we shall find between the $%
\pi _z$ electrons of interest here. The bonding and antibonding $\pi $ bands
have a gapless bandstructure\cite{PhysPropsNanotubesSaitoDresselh}. The
energy near the K points where the bands touch is given by $\varepsilon
^{(1,2)}(\vec{p})=\mp \hbar v_0\left| \vec{p}\right| $ where $\vec{p}$ is
the 2D crystal momentum measured from a $K$ point, and $v_0\,$is a
characteristic velocity (about $5.7\times 10^5\,m/s$ for graphene). From %
perturbation theory within a Wannier description
\cite{NonUnivvdW,JFDUnpubGeneral}, the zero-temperature density-density response $\chi _{KS}$
of independent $\pi _z\,$electrons moving in the groundstate Kohn-Sham
potential of a gapless $\pi $-layer is then of the form $\chi _{KS}(\vec{q},%
\vec{0},\vec{0},z,z^{\prime },\omega =iu)=S(q,z)S(q,z^{\prime })^{*}\bar{\chi%
}_0(\vec{q},iu),$ with $\int Sdz\rightarrow 1\,$as $\vec{q}\rightarrow \vec{0%
}$.
We found \cite{NonUnivvdW,JFDUnpubGeneral} the effective 2D response
$\bar{\chi}_0$ at small $q$ and imaginary frequency $\omega =iu$ to be
\begin{equation}
\bar{\chi}_0(\vec{q},iu)\approx -2\hbar v_0q\left( 1+u^2/(v_0q)^2\right)
^{-1/2}\,\,\, , \label{ChiKS2DNoCutoff}
\end{equation}
consistent with previous real-$\omega$
results\cite{GrapheneRespMarginalFermiLiqu}.
Here we
treat the response in each sheet as strictly two-dimensional, and ignore
certain local-field effects, so that the only consequence of the periodic
potential is to replace the 2D free-electron bare response $-n_0q^2/(mu^2)$
by the zero-gap Bloch response (\ref{ChiKS2DNoCutoff}). This is
justified esewhere \cite
{NonUnivvdW,JFDUnpubGeneral}.
We consider electron density perturbations of form $n_1\exp (i%
\vec{q}.\vec{r}+ut)$ in layer \#1, where $\vec{q}\,$and $\vec{r}$ are
two-dimensional. Such charge disturbances interact via a Fourier transformed
bare coulomb potential, 
\begin{equation}
w_{11\lambda }(q)=2\pi \lambda e^2q^{-1},\,\,\,\,\,\,w_{12\lambda }(q)=2\pi
\lambda e^2q^{-1}\exp (-qD),  \label{CoulombPots}
\end{equation}
for interactions within a layer and between two layers distant $D$,
respectively. Then the RPA equation for the interacting density fluctuation
in layer \#1 as driven by an external potential $v_1^{ext}\exp (i\vec{q}.%
\vec{r}+ut)$ is of time-dependent mean-field form, $n_1=\bar{\chi}%
_0(q,iu)\left( v_1^{ext}+w_{11}n_1\right) $. This applies in the absence of
layer \#2 or equivalently for $D\rightarrow \infty $. Solving for $n_1$we
find a single-layer density-density response 
\begin{equation}
\chi _{11\lambda ,D\rightarrow \infty }\equiv n_1/v_1^{ext}=\bar{\chi}%
_0/(1-w_{11\lambda }\bar{\chi}_0).  \label{Chi11DInf}
\end{equation}
With two layers present, the density response obeys coupled RPA equations $%
n_1=\chi _{11\lambda ,D\rightarrow \infty }(v_1^{ext}+w_{12\lambda }n_2)$, $%
n_2=\chi _{22\lambda ,D\rightarrow \infty }(v_2^{ext}+w_{21\lambda }n_1)$.
The solution is $\vec{n}=\mathbf{\chi }\vec{v}^{ext}$ where $\vec{n}%
=(n_1,n_2)^T$ and similarly for $\vec{v}^{ext},$ while the components of the $%
2\times 2$ matrix $\mathbf{\chi }$ are $\chi _{11\lambda ,D}=\chi
_{11\lambda ,D\rightarrow \infty }/(1-w_{12\lambda }\chi _{11\lambda
,D\rightarrow \infty })$ and $\chi _{12\lambda ,D}=w_{12\lambda }\chi
_{11\lambda ,D\rightarrow \infty }\chi _{11\lambda ,D}$. For the case of two
identical layers considered here, the other elements are $\chi _{22\lambda
,D}=\chi _{11\lambda ,D}$ and $\chi _{21\lambda ,D}=\chi _{12\lambda ,D}$.

In the present system the response (\ref{Chi11DInf}) of a single layer,
continued to the real frequency axis, yields no lightly damped plasmons
(poles) for small $q$, so that a sum of plasmon zero-point energies cannot
be used to evaluate the vdW interaction. Instead we consider the
electromagnetically non-retarded groundstate electronic correlation energy,
which for a general inhomogeneous electronic system is given exactly by the
adiabatic connection fluctuation-dissipation theorem (see e.g.
\cite{JFDIJQC04}):
\begin{equation}
E_c=-\frac \hbar {2\pi }\int_0^1d\lambda \int d\vec{r}d\vec{r}\,^{\prime }%
\frac{e^2}{\left| \vec{r}-\vec{r}\,^{\prime }\right| }\int_0^\infty \Delta
\chi _\lambda (\vec{r},\vec{r}\,^{\prime },iu)du.  \label{GenACFFDT}
\end{equation}
Here $\Delta \chi _\lambda =\chi _\lambda -\,\chi _0$, where $\chi _\lambda $
is the electron density-density response function at reduced coulomb
interaction $\lambda e^2/\left| \vec{r}-\vec{r}\,^{\prime }\right| $.
Applying (\ref{GenACFFDT}) to the present layer geometry and Fourier
transforming parallel to the layers we find that the separation-dependent
part $E^{vdW}/A\,$of the energy per unit area is: 
\begin{eqnarray}
&&\frac {E_c(D)-E_c(\infty )}{A}=-\frac \hbar \pi \int_0^\infty du\int_0^1%
\frac{d\lambda }\lambda \int_0^\infty \frac{2\pi qdq}{(2\pi )^2}  \nonumber
\\
&\times& \left(w_{11\lambda} \left( \chi _{11\lambda D} -\chi _{11\lambda
,D\rightarrow\infty } \right) +w_{12\lambda } \chi _{12\lambda D} \right) .
\label{EvdWOnA}
\end{eqnarray}
Within the RPA\ approximation, Eqs. (\ref{ChiKS2DNoCutoff}) and (\ref
{CoulombPots}), plus (\ref{Chi11DInf}) and the equations following it, show
that each term of form $w\chi \,$in (\ref{EvdWOnA}) depends on $u$ solely
through the dimensionless combination $x=u/(v_0q)$. The remaining dependence
of $w\chi $ on $q$ is solely via $y=qD.$ Thus (\ref{EvdWOnA}) has a scaling
form 
\begin{eqnarray}
E^{vdW}/A &=&\hbar \int_0^1\frac{d\lambda }\lambda \int_0^\infty
qdq\int_0^\infty duG(\lambda ,\frac u{v_0q},qD)  \nonumber \\
&=&\frac{\hbar v_0}{D^3}\int_0^1\frac{d\lambda }\lambda \int_0^\infty
y^2dy\int_0^\infty dxG(\lambda ,x,y)\,  \label{EGrapheneScalingForm}
\end{eqnarray}
where $G(\lambda ,x,y)\,\,$is independent of $D$. We numerically evaluated
the dimensionless 3D integral in (\ref{EGrapheneScalingForm}) for graphene
parameters\cite{NonUnivvdW,JFDUnpubGeneral}, giving the interaction
energy per unit area in Gaussian esu units: 
\begin{equation}
E^{vdW}/A=-7.745_7\times 10^{-2}\hbar v_0D^{-3}=-2.003_6\times
10^{-2}e^2D^{-3}  \label{ECrossEqBOnD3}
\end{equation}

This $D^{-3}\,$form shows that the gapless $\pi $-conjugated planes behave
in this respect more like metals ($E\propto -D^{-5/2}$) than insulators ($%
E\propto -D^{-4}$) , despite the lack of undamped 2D plasmon modes on a
single $\pi $ sheet. 

\emph{D. Parallel metallic and }$\pi $\emph{-conjugated planes }

Another interesting case is the interaction between a $\pi$-conjugated layer
and a metallic 2D layer with fermi energy $\varepsilon_F$
(e.g. an undoped and a doped graphene sheet).
For $D>>D_{0}=\hbar
^{2}v_{0}^{2}/(2\pi e^{2}\varepsilon _{F})$ (=$O(1\,nm)$ for $\varepsilon_F
 =O(0.02\,eV)$ as in a bulk graphite layer) the methods described
above give an energy per unit
area (c.f. (\ref{ECrossEqBOnD3})) 
\begin{equation}
E^{vdW} / A\approx -Ce^{2}D^{-3}\ln (D/D_{0})\,\,\,\,\,\,(C\,\, \text{%
constant}).  \label{Graphene2DEGInteraction}
\end{equation}
As in the case of two non-metallic gapless $\pi$ layers, the result (\ref
{Graphene2DEGInteraction}) disagrees with standard theories.

\emph{E. Interaction energy near overlap}
We now discuss possible difficulties with the non-asymptotic,
near-equilibrium energetics of the present systems, especially graphenes.
 The commonest ab initio approach, the LDA, misses distant dispersion
 interactions entirely \cite{KohnMeirMakarovVdWPRL98}, and yet gives a good lattice spacing
\cite{CharlierGonzeMichLDABSGraphite91} and good breathing phonon frequencies
\cite{GrPhonons} in graphite, (unlike GGAs
\cite{DFTShortRangeBeyondRPAYanPrdwKrth,HardMNosSoftMattRydbergEtal}).
Recent experiments
\cite{BenedictMeasGraphiteLayerAttr,CohEnGraphiteThermDesorb+Hertel}
, however, lead one to suspect\cite{GraphiteBindingSemiEmpirHasegawa+04} that
 the LDA pays for its neglect of dispersion physics by severely
 underestimating the equilibrium binding energy of graphite. LDA also
 underbinds related fullerene systems\cite{OrientationC60sTournusCharlier05}.
  This phenomenon has been
 investigated in layered jellium analogs, via fully nonlocal many-body
 correlation theory (\cite{JFDEVsDJellSlabs04}, Fig. 4 of
 \cite{JeilJFDPabloRex04}).  It was found that either layer-layer
 forces or binding energy have serious errors
 near equilibrium, when distant dispersion forces are underestimated.   The
 underestimation is related in turn to the lack of distant correlated
 fluctuations, especially those oriented parallel to the layers.
Thus these low-q fluctuations can have effects even near the equilibrium spacing.
 Addition of explicit $R^{-6}$ vdW terms has been a common remedy for
 stretched graphitic systems
 \cite{UnivGraphiticPotlGirifalco,GraphiteBindingSemiEmpirHasegawa+04},
 and recently  several seamless vdW schemes have been proposed
\cite{DobsonWangPRL99,HardNosSoftMattRevisRydberg+,vdWFnalGenGeomDionPRL04},
  based on approximations for response functions. 
 Refs \cite{HardNosSoftMattRevisRydberg+} and \cite{vdWFnalGenGeomDionPRL04}
  are the most practical, and are qualitatively  successful in graphitics.
  \cite{vdWFnalGenGeomDionPRL04} overestimates the binding energy
  of
 small systems but correspondingly obtains a large binding energy of two
 graphene layers (more than twice that from
  \cite{HardNosSoftMattRevisRydberg+} or from LDA,
  and consistent with experiment).
   Ideally a single theory should give reliable results for small and extended
   systems.   Could it be that the key is a correct treatment of the
fluctuations parallel to a long axis
   in the extended cases, the same fluctuations responsible for
   the unusual asymptotics exposed here that is absent in
   \cite{HardNosSoftMattRevisRydberg+,vdWFnalGenGeomDionPRL04}?  These
   fluctuations are of
   course dominant only at large separations but they might not be negligible
    near the equilibrium spacing, where all wavelengths can contribute.   We
  speculate further that the same physics might apply in other large
  finite $\pi$-conjugated systems (e.g. planar melanin layers, carotenes,
  fullerenes
\cite{OrientationC60sTournusCharlier05}) where, as the system size increases,
 the electronic gap diminishes while longer-wavelength excitations become
 possible.

\emph{Summary and Discussion. }Our new results (see (\ref{WireWirePlasmon}),(%
\ref{ECrossEqBOnD3}),(\ref{Graphene2DEGInteraction}) and Table 1)
 show that that usual sum of $C_{6}R^{-6}$ terms incorrectly predicts the
 dependence of the dispersion energy on separation $D$ for a range of systems.
Simple energy functionals presently available all have standard $%
\sum C_{6}R^{-6}$asymptotics. A finite sum of multipole, or triplet and
higher terms will also not reproduce what we have discussed. The standard
asymptotics fails when the component systems (i) are metallic (or have a
zero electronic Bloch bandgap), and (ii) are spatially extented in at least one
dimension, so that long-wavelength (low-$q$) charge fluctuations can occur,
and (iii) are of nanoscopic dimensions in another spatial direction, so that
the electron-electron screening is reduced compared with 3D bulk metallic
systems, leaving a divergent screened polarizability at low frequency and
wavenumber. 
(Thick metal slabs, for example, violate (iii): they have complete screening
and exhibit a conventional power law, $E\propto -D^{-2}$. See e.g. \cite
{JFDEtAlAustJChem02}). Where free low-$q$ plasmons exist, conditions (i) -
(iii) imply that they will be gapless. The same conditions ensure that the
usual spatially local approximation for the dielectric function \cite
{LifshitzMolAttrFrcsSolsJETP56} is invalid.  Our results provide
unequivocal asymptotic benchmarks that are not satisfied by existing
simplified van der Waals energy formulae, because they do not treat
in enough detail the fluctuations along the extended space dimension.
In Section E we have further
motivated the possibility that the same fluctuation physics may be
relevant in the systems considered here, even near their equilibrium
spacing. Investigation of this question requires a seamless
 energy formalism that is fully nonlocal - e.g.  RPA-like theories
\cite{JFDEtAlAustJChem02,DobsonWangPRL99,JeilJFDPabloRex04,JFDIJQC04}. Such
calculations are only now becoming possible for 3D systems \cite
{RPAGreenFnEnergysolidsMiyake+PRB2002,BoronNitrideInRPA+MariniGarciaRubioPreprint05}, with
no converged results available to date for the present zero-gap cases. 
Simplified vdW energy functionals are therefore certainly needed for
routine modelling, and the above considerations suggest that existing
functionals may need further refinement to take explicit account of
large-scale geometry and/or nonlocal entities such as electronic
bandgap\cite{JFDIJQC04}. We note finally that our work predicts novel
differences in the forces between conducting and nonconducting nanotubes
or wires,  that might be directly measurable for low-index nanotubes at
sub-micron separations\cite{JFDAWhiteUnpub}, and that could even affect
self-assembly processses. These
considerations might also affect the analysis of some seminal experiments%
\cite{BenedictMeasGraphiteLayerAttr,CohEnGraphiteThermDesorb+Hertel}
concerning graphitic cohesion, because these relied at some point on theory
involving a sum of $R^{-6}$ contributions.
\noindent \emph{Acknowledgments}.  We thank I. D'Amico, P. Garcia-Gonzalez,
J. Jung, L. Reining, E. Gray and P. Meredith
 for discussions. JFD acknowledges support from Australian
Research Council grant DP0343926, UPV/EHU, Ecole Polytechnique and CNRS, and
the hospitality of AR and Dr. L. Reining. AR was supported by the Network of
Excellence NANOQUANTA (NMP4-CT-2004-500198), and a 2005
Bessel research award of the Humbolt Foundation.

\end{document}